\documentclass[aps,prb,twocolumn,showpacs,amssymb]{revtex4}

\usepackage[dvips]{graphicx}
\usepackage{dcolumn}
\usepackage{bm}
\def\be{\begin{equation}}
\def\en{\end{equation}}

\newcommand{\av}[1]{\langle{#1}\rangle}
\def\gs{\gtrsim}
\def\ls{\lesssim}
\newcommand{\bi}[1]{\mbox{\boldmath$#1$}}

\def\ve{\varepsilon}
\def\bea{\begin{eqnarray}}
\def\ena{\end{eqnarray}}

\usepackage{color}

\begin{document}


\title{Visualization of stacking faults  in fcc crystals 
in plastic deformations}



\affiliation{Department of Physics, Kyoto University, Kyoto 606-8502,
Japan}

\author{Takeshi Kawasaki and Akira Onuki}
\affiliation{Department of Physics, Kyoto University, Kyoto 606-8502,
Japan}


\date{\today}

\begin{abstract}
Using  molecular dynamics simulation, 
we investigate the dynamics of 
stacking faults in fcc crystals 
in uniaxial stretching in 
 a Lennard-Jones  binary mixture 
composed of 4096 particles  in three dimensions.   
We visualize   stacking faults using 
  a disorder variable $D_j(t)$ 
 for  each particle $j$ 
 constructed from   local bond order parameters 
based on spherical harmonics (Steinhardt order
parameters). 
Also introducing a  method of bond breakage, 
 we examine how stacking faults 
 are formed  and  removed by collective 
particle motions. These processes are relevant 
in plasticity of fcc crystals. 
\end{abstract}

\pacs{83.10.Rs, 61.72.Nn, 62.20.F-, 61.66.-f}
  

\maketitle


\section{Introduction}

In crystal plasticity, dislocations play a major role 
\cite{Hirth,Abraham,dislocation,KubinD,postmortem}. 
They  tend to appear 
  around  point defects, preexisting 
dislocations,  grain boundaries, and precipitates. 
They then  grow to form   slip planes, which extend 
  over large spatial regions  
under the strong influence of the crystal structure 
\cite{Yama,Swy,Ma,Li,Hatano,Shimokawa,twin}. 
In polycrystal,  birth and growth of 
dislocations are strongly influenced by 
grain boundaries, so  the grain size 
is a key factor of  plasticity. 
With decreasing the grain size, 
sliding  motions of the particles at 
grain boundaries become increasingly 
important \cite{Hahn}, as  has been revealed by 
 molecular dynamics simulations 
\cite{Yip,Yama,Swy1,Jack,HamanakaShear,Gerlich}. 
In crystal and polycrystal, 
 plastic events have been observed  as  
bursts or avalanches  
spanning wide ranges of space and time 
scales. We may mention experiments of 
acoustic emission  \cite{dislocation} 
and  transmission 
electron microscopy \cite{postmortem}.

In particular, fcc crystals may be regarded  as a  
  sequence  of  
 closed-packed $\{111\}$ layers\cite{Hirth}  
 symbolically represented by $ABCABCABC$, 
 while hcp  crystals are represented by 
   $ABABAB$. If the free energy of the fcc structure 
   is slightly lower than that of the hcp structure, 
 the regular fcc sequence is often violated by  
stacking faults or twin faults.    
For example, an intrinsic  stacking fault 
is represented by  $ABC{{AC}}ABC$,  
where ${B}$ is missing 
between the  middle two-particle layers ${ AC}$. 
This planar defect is produced by 
collective slip  motions $B \to C$, 
$C\to A$, and $A\to B$ ($B \to A$, 
$C\to B$, and $A\to C$) on the right (left) 
side of the defect plane. 
If its size is finite within the crystal, 
a partial dislocation can be found  at  its border  
 with a Burgers vector of type  
 $(a/6)\av{1{\bar 2}1}$, where $a$ is the lattice constant. 
 As is well-known\cite{Hirth}, 
this stacking fault can disappear  with 
 superposition of another 
  partial displacement  of type  
 $(a/6)\av{{\bar 1}{\bar 1}2}$, 
where  a full dislocation of type 
$(a/2)\av{0 {\bar 1} 1}$  remains at its border.  
  Plastic deformations can be 
 achieved  by collective stacking reorganization 
   for metals   with  small grain sizes 
(nanocrystals) \cite{Yama,Swy}. 
Stability of partials depend on various factors 
such as the grain size, 
the applied stain rate, 
and the potential energies for the partial 
displacement. Full dislocation slips 
have rather been observed in metals with 
 large grain sizes and with  
relatively large stacking fault energies.

In colloids, direct observations of 
particle configurations  are 
particularly informative
\cite{Schall,Poon,Weiz,De,Roth,Blaa,Blaa1,Pusey,Dhont,Chaikin,Solomon}. 
If the colloid interaction is  
short-ranged and hard-core-like,  
there is almost 
no free energy difference between the 
fcc and hcp stuctures \cite{Pronk}, 
so fcc and hcp  stacking 
layers   often appear randomly 
as random hexagonal closed-packed (rhcp) states 
\cite{Poon,Weiz,De,Blaa,Blaa1,Pusey,Dhont,Chaikin,Solomon}. 
A pure fcc crystal 
can be realized by sedimentation of colloid particles 
onto a patterned $[100]$  substrate \cite{Blaa}.
It is known that the earth gravity and small
mechanical perturbations  strongly 
affect the structure of colloid crystals \cite{Pusey}.
In oscillatory shear, 
the spatial distribution  of stacking faults 
was measured \cite{Solomon}. Dislocations 
and slip planes  have been observed 
in indentation experiments \cite{Schall,Roth}.

In three dimensions,   
disordered crystal states are 
very  complex under the strong influence of  the 
crystal structure. To analyze them, 
Steinhardt  {\it et al.}\cite{Steinhardt} 
 introduced   tensorial bond-order parameters 
based on the spherical harmonics.  
With their  method in analysis, 
the crystallization process 
has been simulated 
\cite{tenWolde,Auer,Dellago,Koplik,Snook,KawasakiTanaka3d}. 
Stacking faults have been observed in 
growing crystal domains \cite{tenWolde,Auer,Koplik,Snook}. 
It has also been used to examine  
the structural heterogeneity in glass 
\cite{KawasakiTanaka3d,KawasakiJCP}. 
In this paper,  we will present 
results of  molecular dynamics simulation 
on the fundamental dynamical processes  
 of  stacking faults in uniaxial stretching. 
In this problem, it is highly desirable  
to  develop   the  method of visualizing 
 these collective and complex phenomena. 
   With this purpose,  we will introduce 
  a disorder variable $D_j$   constructed from 
the Steinhardt  order parameters 
\cite{KawasakiJCP}, which represents the deviation of 
hexagonal crystalline order  for each particle $j$. 
In our simulation,  we will domonstrate that 
   appearance and disappearnce 
 of intrinsic stacking faults  give rise to crystal plasticity.

The organization of this paper is as follows.
In Sec.II,    our simulation method will be explained, 
 where a  method of bond breakage and  the  disorder variable $D_j(t)$ 
 will be  introduced. In Sec.III,  
  we will  visualize  stacking faults  
on the basis of our simulation 
to understand the large-scale 
configuration changes in plastic flow.  
We will  examine rapid time-evolution 
of the potential energies 
and the  displacements 
of the particles close to 
a stacking fault at  plastic events. 
We shall see that the edge of an expanding stacking fault 
propagates with  a velocity close to the transverse sound velocity, 
while the particle velocities 
at the slip plane are much slower.  
We will also display the averages 
of various physical quantities over all the particles,  
including the stress-strain curve, on long time scales. 
They exhibit intermittent changes upon plastic events.

\section{Background of  Simulation} 
\subsection{Model and simulation method}

We treat three-dimensional   binary mixtures 
composed of two atomic species 1 and 2. 
The total particle number is $N=N_1+N_2=4096$. 
The composition  of 
the large particles is written as $c=N_2/N$. 
The  particles  interact  via  
truncated Lennard-Jones (LJ) potentials,  
\begin{equation}
v_{\alpha\beta} (r) = 
4\epsilon \left[ \left(\frac{\sigma_{\alpha\beta}}{r}\right)^{12} - 
\left(\frac{\sigma_{\alpha\beta}}{r}\right)^6\right] -C_{\alpha\beta} ,
\label{eq:LJP}
\end{equation}
which are characterized by the energy $\epsilon$ 
and the interaction lengths    
$\sigma_{\alpha\beta} = 
(\sigma_\alpha +\sigma_\beta )/2$ $(\alpha,\beta=1,2)$. 
Here $\sigma_1$ and $\sigma_2$ are  
 the  soft-core diameters of the two species. 
For $r>r_{\scriptsize{\textrm{cut}}} 
=2.25\sigma_2$, we set $v_{\alpha\beta} =0$
and the constant $C_{\alpha\beta}$ 
ensures the continuity of $v_{\alpha\beta}$
at $r=r_{\scriptsize{\textrm{cut}}}$. 
The  mass ratio is fixed at  
$m_1/m_2 = (\sigma_1/\sigma_2)^3$.
Initially,  the particles were  in a cubic box, whose 
length $L_0$ was  
determined by 
$L_0^3= N_1\sigma_1^3 +N_2\sigma_2^3$.
For $c=1$, we set $L_0= 15.85\sigma_2$.  
The time step of integration is $0.005\tau_0$ with
\be 
\tau_0 = \sigma_2 \sqrt{m_2/\epsilon}.
\en 

After preparation of the initial particle 
configurations  at $T=0.015\epsilon/k_B$, 
we  applied uniaxial strain along the $z$ axis for  $t>0$. 
To this end, we moved  the position of the top plate 
of the cell  $z=L(t)$ as  
\be 
L(t)= L_0(1 + \dot{\ve}t ),
\en 
where  $\dot{\ve}$ is the elongation rate 
chosen to be      
\be 
\dot{\varepsilon }=3.0\times 10^{-5}\tau_0^{-1}.
\en  
All the side walls parallel to the $z$ axis 
 were deformable  free boundaries\cite{Falk3D}. This free-surface condition 
 was  possible due to the attractive parts 
of the LJ potentials.    As in our 
two-dimensional simulation \cite{Shiba}, 
we  divided the cell 
into  a bottom  layer 
in the region $0<z< d_0= 0.0625L_0$,  
a top layer in the region $L(t)-d_0 <z< L(t)$,  
and an interior region in the region $d_0<z< L(t)-d_0$.  
Each layer contained   pinning centers  fixed at 
 positions ${\bi R}_j$ ($j=1,\cdots, N_b$). 
 At each pinning center, a  particle at position ${\bi r}_j$ 
 is bound by a  spring potential,    
\begin{equation}
u_{j} (\bm{r}_j -\bm{R}_j) 
=  K_s  |\bm{r}_j  -\bm{R}_j|^2/2, 
\label{eq:SPP1}
\end{equation} 
where we set $K_s = {20\epsilon}{\sigma_2^{-2}} $.
In this paper, the  number of the bound particles 
in each layer is chosen to be  $N_b=256$.  
Then the  number of the unbound particles 
in the interior is $N_{\rm ub}= N-2N_b= 3584$. 
The positions of the pinning centers were  
independent of time in the bottom layer, 
but depended   on time  in the top layer  as 
$
\bm{R}_j (t)= \bm{R}_j(0)+ L\dot{\varepsilon }t {\bi e}_z,
$ where ${\bi e}_z$ is the unit vector in the $z$ axis. 
The bound  particles in the layers 
interacted  with the other unbound and bound particles  
via the LJ potentials in Eq.(1). 
The total potential energy 
 consists of two parts  as 
\be 
U_{\rm tot}= \frac{1}{2}
\sum_{j,k{\in {\rm all}} }v_{\alpha\beta}(|{\bi r}_j-{\bi r}_k|) 
+ \sum_{j\in {\rm layer}}u_{j}(|{\bi r}_j-{\bi R}_j|), 
\en 
where the summation in the first term 
is over all the particle pairs and 
that in the second term is over the $2N_b$ bound particles.

The $N_{\rm ub}$ unbound particles were   governed by 
the usual Newton equations. 
Their  positions ${\bm{r}}_j$ obeyed      
\be 
m_\alpha \ddot{\bm{r}}_j =  
- \frac{\partial }{\partial\bm{r}_j}U_{\rm tot},
\en 
where $\ddot{\bi r}_j= d^2{\bi r}_j/dt^2$. 
In our simulation,  the unbound particles 
in touch with the bound particles turned out to be 
clamped to the boundary  layers 
without slip and detachment during stretching. 
 They also did not  penetrated   into 
the layers deeper than  $\sigma_2$.  
On the other hand, we attached 
  a Nos\'e-Hoover thermostat \cite{Shiba,nose} 
 to each boundary layer.  
 That is, we introduced  
  two thermostat  variables   $\zeta_{\rm bot}(t) $ and 
$\zeta_{\rm top}(t) $.  Let ${\cal B}$ 
represent   the top or bottom layer.  
The bound particles $j \in {\cal B}$ obeyed 
\be 
m_\alpha \ddot{\bm{r}}_j =  
- \frac{\partial }{\partial\bm{r}_j} U_{\rm tot} - 
\zeta_{\cal B}  m_\alpha (\dot{\bi r}_j-{\bi v}_{\cal B}), 
\en 
where   $\dot{\bi r}_j= d{\bi r}_j/dt$,      
and ${\bi v}_{\cal B}$ is the boundary velocity 
equal to $L\dot{\varepsilon }{\bi e}_z$ at the top 
and to  ${\bf 0}$  at the  bottom. 
The  thermostat  variables $\zeta_{\cal B}$  were governed by        
\be
{\tau_{\rm{NH}}^2}\frac{d}{dt} {\zeta}_{\cal B} =  
 \frac{1}{3 N_bk_BT}\sum_{j \in {\cal B} } 
m_\alpha|{\dot{\bi  r}_j-{\bi v}_{\cal B}|^2} -1 , 
\en
where  $\tau_{\rm{NH}}$ is the thermostat 
characteristic time.  In this paper, 
we  used a very short time 
$\tau_{\rm{NH}}= 0.072\tau_0$. 
Then the top and bottom 
layers  played  the role of   efficient 
 thermostats subtracting  
 the extra energy released at plastic deformations  
in the bulk. 
As a result, the local temperature 
(the  particle kinetic energy averaged in a narrow 
region) became nearly homogeneous except for short 
duration periods of plastic events  in the bulk region.

We explain our 
preparation method of 
 the  initial particle configurations \cite{Falk3D,KawasakiJCP}. 
We followed four steps in the time region  $t<0$   
by  solving  Eqs.(6) and (7) 
with ${\bi v}_B={\bi 0}$. 
In the first three steps, we imposed 
 the periodic boundary condition 
in all the directions with $N=4096$. 
(i) First, we created  
random particle configurations in a liquid state 
at   $T=1.75\epsilon/k_B$ in a 
time interval  longer than  $10^3\tau_0$. 
(ii) Second,  we  quenched the system to 
$T=0.55\epsilon/k_B$  below 
the melting  and equilibrated it 
in a time interval longer than  $10^4\tau_0$. 
In this intermediate quenching,  
 crystalline configurations were realize \cite{Falk3D}. 
(iii) Third,  we further quenched the system  to 
the final temperature $T=0.015\epsilon/k_B$ 
and annealed it in a time interval longer than  $10^3\tau_0$. 
There were  almost no structural changes in this step. 
(iv) Fourth,  the periodic  boundary 
condition along the $x,y,z$ axises  were changed such that 
 the boundaries perpendicular 
to  each axis  became  free surfaces. 
This change  instantaneously caused 
a small decompression and   an expansion 
along each axis by about $2\%$.  
We then  annealed the system 
 for a time  longer  than $10^3\tau_0$ with $T$  held 
 fixed. At this low temperature,     
 there was no  detachment of the particles from 
 the free surfaces due to the attractive interaction.   
 After this procedure,
we chose the particle positions in the boundary layers as
the initial pinning points ${\bf R}_j $  and introduced the
spring potential in Eq.(5) 
between  the bound particles and the pinning centers  
 and the  Nos\'e-Hoover thermostats of the boundary layers.
We set  $t=0$ after these procedures.   

\subsection{Bond breakage, disorder variable, 
and stress}

In dense particle systems,  
configuration changes  
can be visualized 
with  the method of bond breakage \cite{yo1}. 
For each particle 
 configuration  at a time $t$, 
a pair of particles $i\in\alpha$ and 
$j\in\beta$ is considered to be bonded if
\begin{equation}
r_{ij}(t) = |\bm{r}_i(t)-\bm{r}_j (t)| \le A_1 
\sigma_{\alpha\beta}.
\end{equation}
We set $A_1=1.3$, 
for which  $A_1\sigma_{\alpha\beta}$ 
is  slightly longer  than the first peak 
distance  of the pair correlation 
function $g_{\alpha\beta}(r)$. 
For our fcc crystals for $c=1$, 
the corner-face distance is equal to 
$1.088\sigma_2$ (which is slightly shorter than 
the minimum distance $2^{1/6}
\sigma_2=1.122\sigma_2$ of the Lennard-Jones potential). 
The corner-corner  distance 
or the lattice constant 
is then $a=1.088\times 2^{1/2}\sigma_2=1.539\sigma_2$. 
Thus, for $A_1=1.3$, the 12 nearest 
particle pairs at corner-face 
and face-face positions 
in a unit cell are bonded and  the pairs at corner-corner 
positions are not bonded. 
 After a time interval $\Delta  t$, 
the bond is regarded to be broken if 
\begin{equation}
r_{ij}(t + \Delta t)\ge A_2\sigma_{\alpha\beta}.   
\label{eq:BBOD2}
\end{equation} 
This definition of bond breakage  
is insensitive to 
the value of $A_2$ as long as $A_2 $ is larger 
than $A_1$ and $A_2\sigma_{\alpha\beta}$ is shorter than the 
second peak position  of 
$g_{\alpha\beta}(r)$  \cite{yo1}. 
In our simulation, 
the particle pairs 
with  a broken bond 
in this definition underwent 
 a relative motion 
 mostly of order  $\sigma_{\alpha\beta}$. 
The broken bond number  
in  time interval $[t, t+\Delta t]$ 
will be denoted by  $\Delta N_b(t)$ 
as a function of time $t$
at  fixed $\Delta t$. The number of the particles with 
broken  bonds in this time interval 
is twice of  $\Delta N_b(t)$.

Next, to examine  the structural order, 
 we  introduce Steinhardt  
order parameters 
\cite{Steinhardt, tenWolde,Auer,Dellago,Koplik,Snook,KawasakiTanaka3d},  
\bea
q_{\ell m}^{j}(t)=\frac{1}{n_b^{j}(t)}
\sum_{k \in {\rm bond}}Y_{\ell m}({\bm r}_{jk}(t)),
\label{eq:Alpha}
\ena
which are defined for each unbound particle $j$. Here,  
$Y_{\ell  m}({\bi r})$ are the  spherical harmonics functions  
of degree $\ell$  with $-\ell \le m\le \ell$,    
 depending   on the direction of a vector $\bi r$. 
In our system, the local 
crystal structures are mostly fcc and  
we choose  $\ell=6$.  
 In  Eq.(12), 
 we set ${\bi r}_{jk}(t)  =\bm{r}_j(t) -\bm{r}_k(t)$,  
where the particle $j\in \alpha $ at position $\bi{r}_j(t)$ 
and    the particle $k\in \beta$ 
at position $\bi{r}_k(t)$  are bonded 
($r_{jk} <A_1 \sigma_{\alpha\beta}$). 
The number of the  particles bonded to the particle $j$ 
is written as $n_b^j(t)$. 
This criterion of bonded particles 
is the same as  in Eq.(10), so 
$n_b^j(t)$ is mostly equal to 12 in fcc crystal regions.
 We  further introduce a disorder 
variable for each particle $j$ by \cite{KawasakiJCP} 
\be
D_j(t) = \frac{1}{n_b^{j}(t)}
\sum_{k \in {\rm bond} }~ \sum_{m=-6}^{6} 
|q_{6m}^{j}(t)-q_{6m}^{k}(t)|^2, 
\en
where the average over the bonded particles 
$j$ is taken.  For particles around defects $D_j(t)$  is 
in the range $[0.1,0.7]$, while 
for a perfect crystal 
 $D_j(t)$  is nearly zero at low $T$ 
 ($D_j\ls 0.005$ in fcc regions for $c=1$). 
Thus   $D_j(t)$ represents  
  the deviation from 
hexagonal crystalline order  for each particle $j$. 
It is convenient 
to introduce   the average of $D_j(t) $ over the $N_{\rm ub}$ 
unbound particles ($j \in {\rm ub}$),  
\be 
{\bar D}(t)=\sum_{j\in {\rm ub}} D_j(t)/N_{\rm ub}. 
\en 
which represents the overall degree of disorder. 
We have originally devised a disorder variable 
in two dimensions \cite{Hama}.

During stretching, 
we also calculated    the average of 
the $zz$ component of the  stress 
 \cite{stress-comment}. It is expressed as 
summations  over the unbound particles in the form,   
\be 
\av{\sigma_{zz}}(t)= \frac{-1}{V}\bigg[
\sum_{j\in {\rm ub}} m_\alpha \dot{z}_j^2
-\sum_{jk\in {\rm ub}}\frac{z_{jk}^2}{2r_{jk}}v_{\alpha\beta}'(r_{jk}) 
\bigg],
\label{eq:Stress}
\en 
where $(x_j,y_j,z_j)$ are 
 the particle positions, $
 (\dot{x}_j,\dot{y}_j,\dot{z}_j)$ 
 are  the velocities,  
   $\bm{r}_{jk} =  \bm{r}_j-\bm{r}_k 
=({x}_{jk},{y}_{jk},{z}_{jk})$,  
are the relative position vectors with  $r_{jk}=
 |\bm{r}_{jk}|$,   and $ v_{\alpha\beta}(r_{jk})$  are 
 the pair potentials with 
 $v_{\alpha\beta}'(r)= 
d v_{\alpha\beta}(r)/d r$. 
Here $V$ is the volume in the bulk interior 
region, but we  set  $V=(L_0-2d_0)L_0^2$ 
neglecting its weak time-dependence. 
We expect that 
$\av{\sigma_{zz}}(t)$ is approximately equal to 
the stress acting on the top and bottom 
  after smoothing out 
of the contributions 
from the rapid spring motions  
 \cite{Shiba}.    
To be precise, the overall stress distribution 
 is somewhat complicated in our system 
 due to  the clamping condition at the top and 
bottom and the free-boundary condition on the side walls. 
See  inward deformations of the side boundaries 
at strain $\sim 0.06$ in Fig.5 below.

\section{Simulation  results }

\begin{figure}[t]
\includegraphics[scale=0.45]{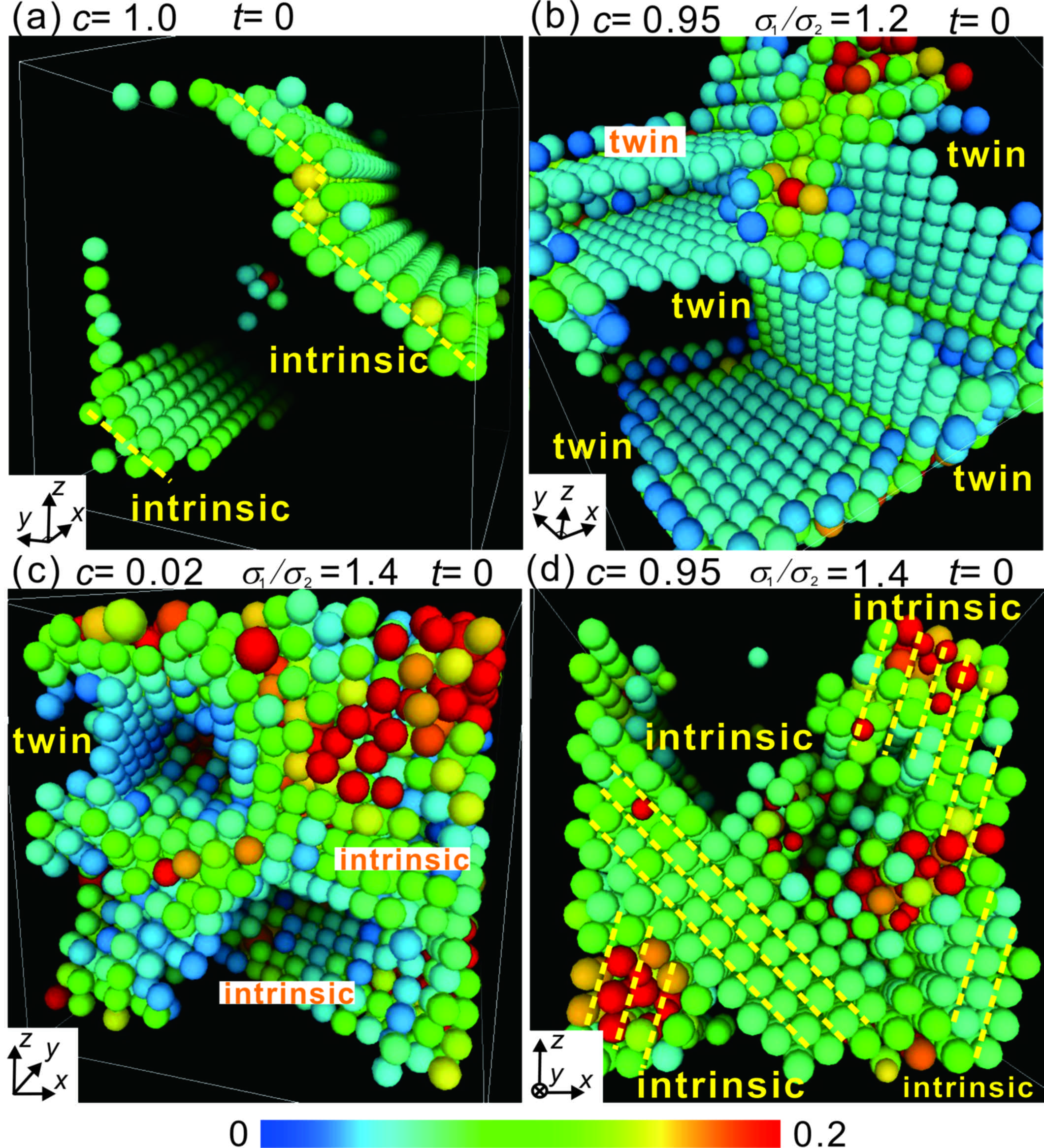}
\caption{(Color online) 
Relatively disordered  particles  
with $D_{j}>D_0$ forming stacking faults in 
fcc crystals at $t=0$, where we set 
$(c, \sigma_2/\sigma_1,D_0)$  equal 
to 
(a) $(1, 1.2,0.05)$, 
(b) $(0.95, 1.2, 0.03)$, 
 (c) $(0.02, 1.4, 0.03)$,  and 
 (d) $(0.95, 1.4, 0.05)$. 
   Particles   are written as spheres with 
  diameter $\sigma_{1}$  or $\sigma_{2}$.  
Their   colors represent   $D_{j}$ 
according to  the color bar. 
Stacking faults visualized are 
 intrinsic  ones extending 
 throughout the cell in (a), while 
 they are  twin  ones 
 stemming  from 
a grain-boundary-like 
disordered region  in the upper part in (b). 
 Twin  and intrinsic ones are mixed 
 in (c).  Successive intrinsic ones 
 form hcp regions  in (d). 
}
\end{figure}

\begin{figure}[t]
\includegraphics[scale=0.38]{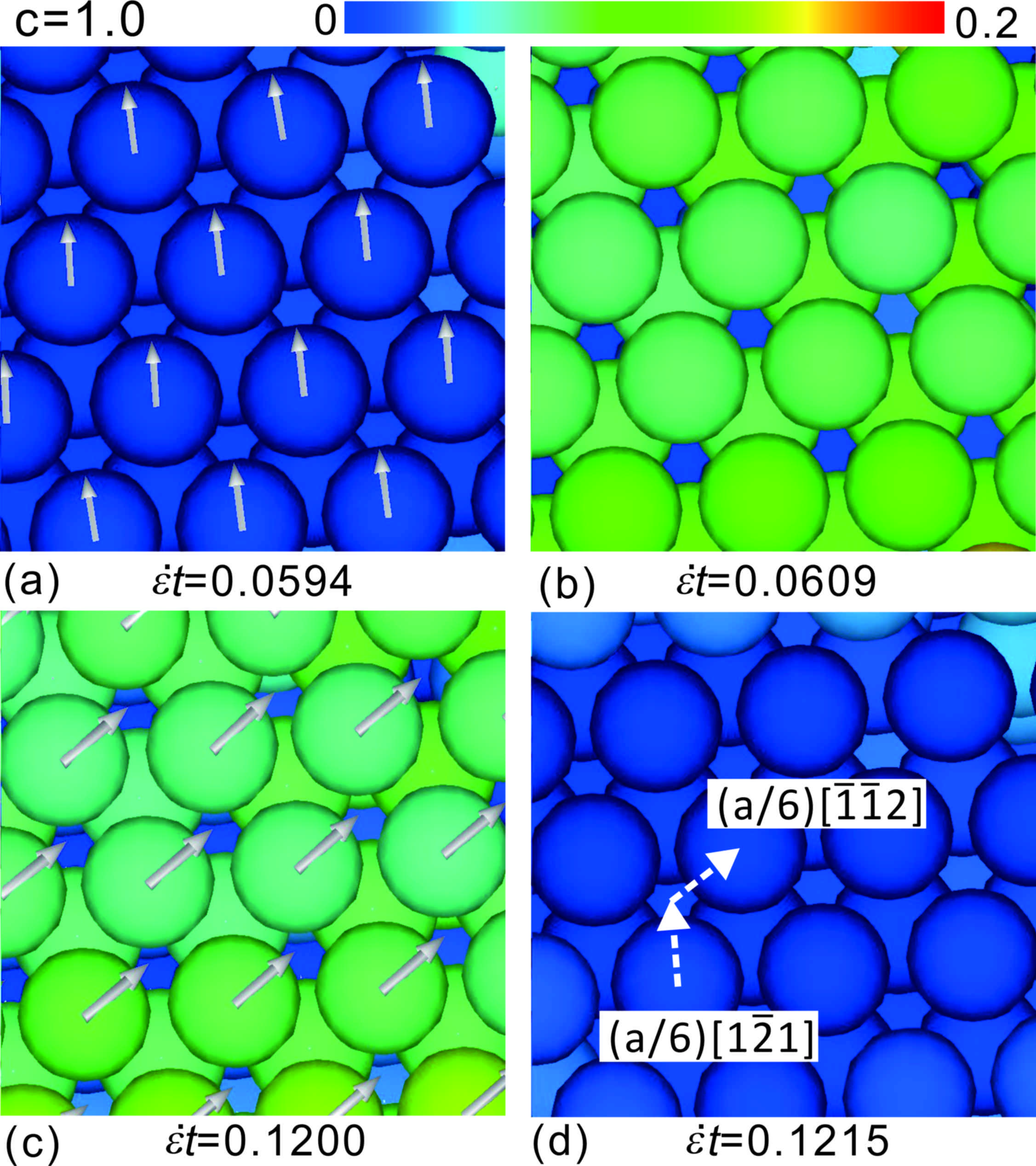}
\caption{(Color online) 
Particles twice undergoing partial relative displacements 
  on the  $(111)$ plane 
for $c=1$ at  $\dot{\varepsilon}t= 0.0594$ in (a) (top left), 
$0.0609$ in (b) (top right),  $0.1200$ in (c) (bottom left), 
and $0.1215$ in (d) (bottom right). 
Particle  colors represent 
$D_j(t)$ according to the color bar. 
In (a) a perfect fcc stacking  is realized. 
Arrows indicate a partial relative 
displacement between the top and next layers 
 given by $0.8 (a/6)[{1} {\bar 2} { 1}]$ here, which 
takes place  just after this moment.  In (b) 
an intrinsic  stacking fault just created  is shown 
  after an elapse of $30\tau_0$, where 
 $D_j$ of   the top and next layers have 
  increased to a higher value (in green).  
 The third layer  remains to have the lowest 
 fcc value of $D_j$  (in blue).  
In (c), after a long elapse of time, 
the same stacking fault is going to disappear with    a second 
partial displacement $(a/6)[{\bar{1} {\bar 1} 2}]$ 
indicated by arrows. In (d), a perfect fcc stacking  is recovered. 
Data  in Fig.1(a)  and Figs.2-7  are from the same simulation run. 
}
\end{figure}

\begin{figure}[t]
\includegraphics[scale=0.43]{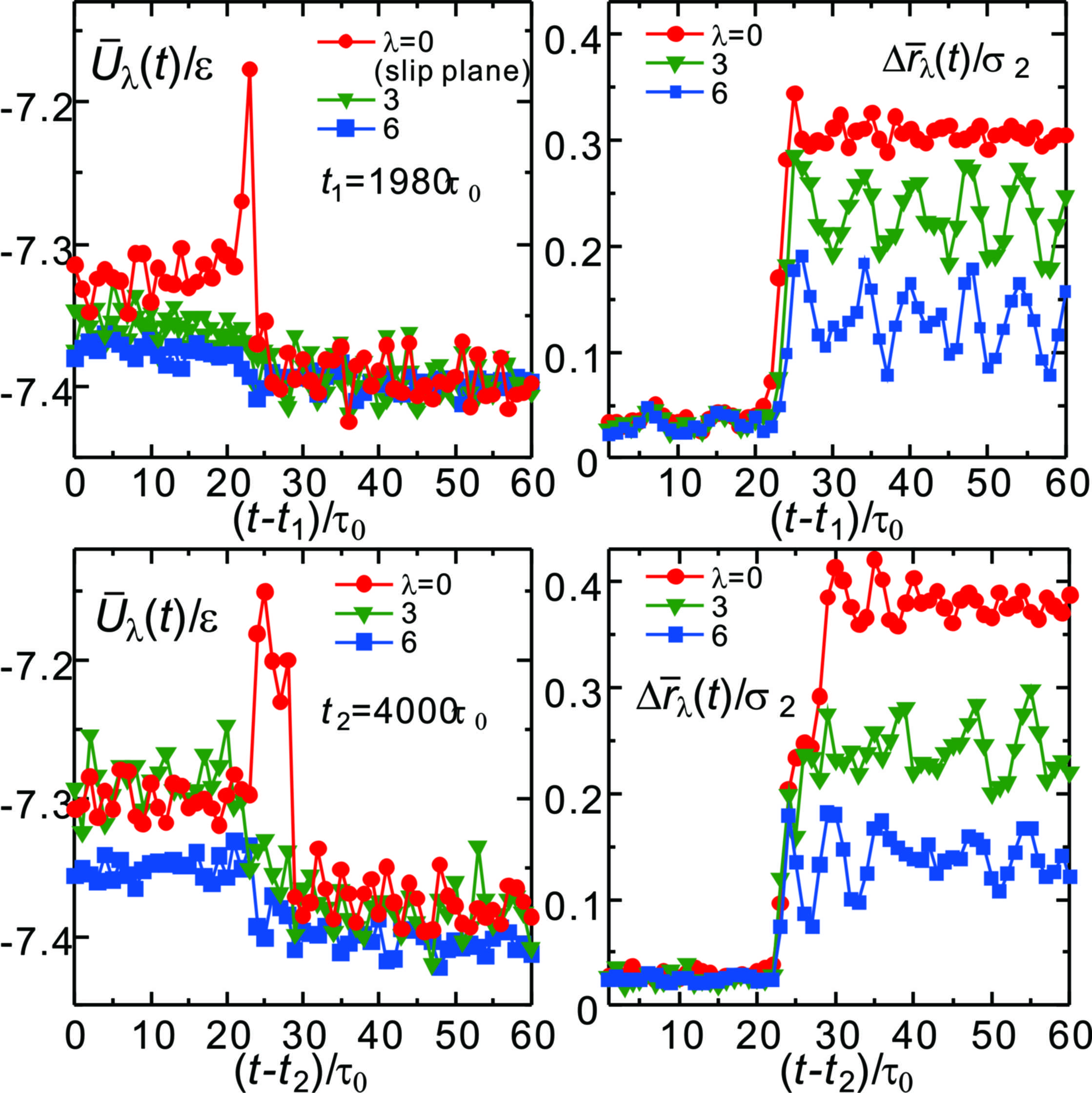}
\caption{(Color online) 
Left: Potential energy  $\bar{U}_\lambda (t)$ 
averaged over 50 particles on  the $\lambda$-th layer 
below the    stacking fault in Fig.2 
 emerging at  $t-t_1\sim 20 \tau_0$ 
with $t_1=1980\tau_0$ (top) and 
disappearing at $t-t_2\sim 20\tau_0$ with $t_2=4000\tau_0$  
(bottom). This stacking fault 
 is  composed of the layers $\lambda=0$ and $-1$. 
Right: Displacement 
$\Delta{\bar{r}}_\lambda(t)$ averaged 
over the same 50 particles  
at the formation  (top)  and  at the 
disappearance (bottom) of the stacking fault. For $\lambda=0$,   
 the maximum velocity ($\sim 0.1\sigma_2/\tau_0$)
  is much slower than the edge  velocity ($
 \sim 4\sigma_2/\tau_0$). }
\end{figure}

\begin{figure}[t]
\includegraphics[scale=0.4]{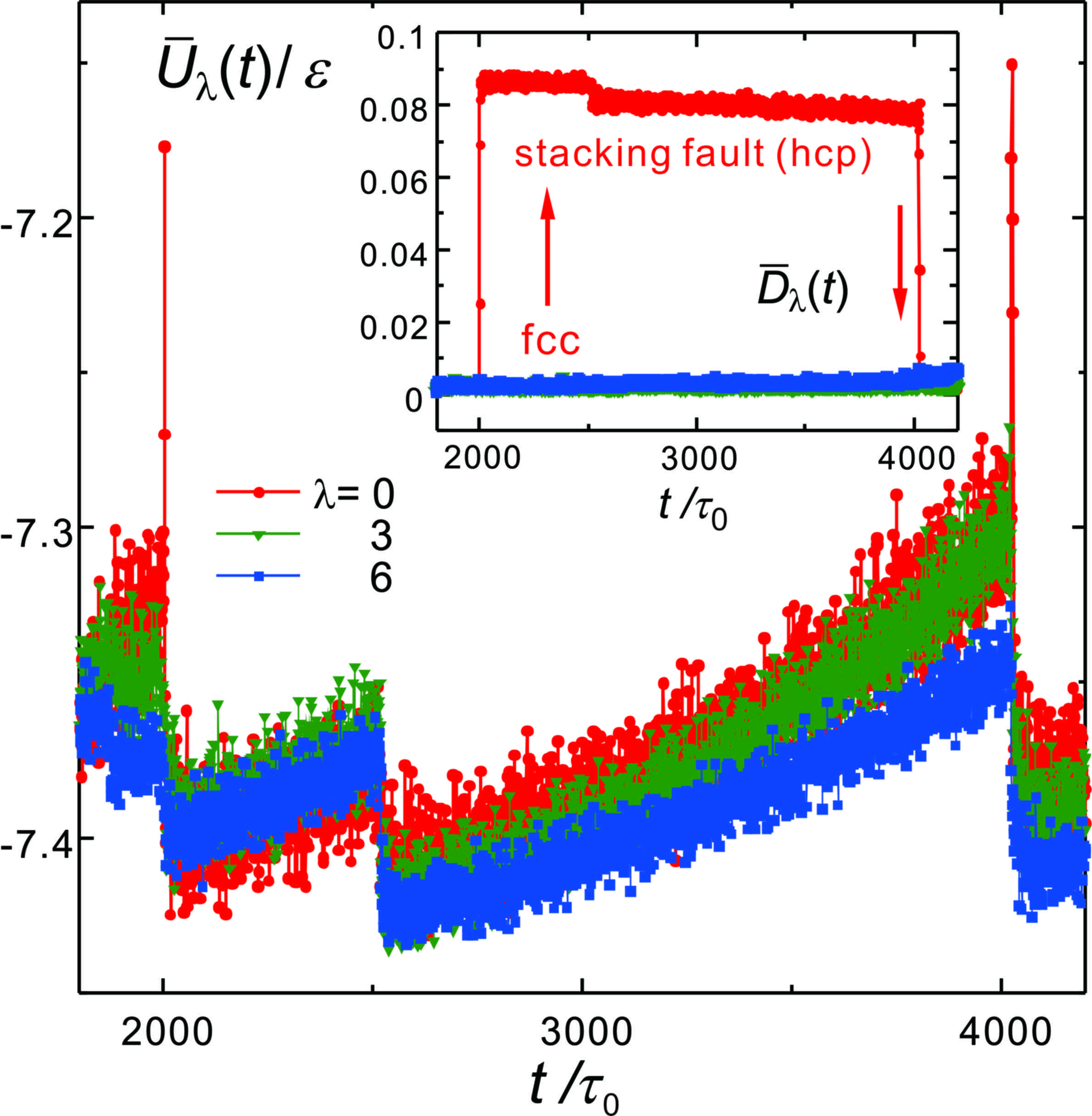}
\caption{(Color online) 
Average potential energy  $\bar{U}_\lambda (t)$ 
 and average disordered variable  
$\bar{D}_\lambda(t)$ (inset)  for 
$\lambda=0, 3$, and $6$ over a long time interval. 
The layer element $\lambda=0$ undergoes 
two partial displacements at 
 $t\sim 2000\tau_0$ and  $4000\tau_0$, 
where $\bar{D}_\lambda(t)$  increases or decreases 
steeply  for $\lambda=0$.  
Another slip takes place at $t\sim 2500\tau_0$ 
apart  from these layer elements, 
at which the elastic energy 
density decrease suddenly for $\lambda=3$ and 6 but does not 
change for $\lambda=0$.   }
\end{figure}

\subsection{Stacking faults at $t=0$}

We mostly realized 
intrinsic  stacking faults 
with sequence $ABC{\bi {AC}}ABC$ 
 at low levels of disorder 
in unstrained states 
and  under   stretching. Furthermore, 
with addition of  structural disorder 
  in  dilute mixtures,  
we also realized twin  faults 
with sequences $ABCAB{\bi C}BACBA$ 
during  crystallization (in the preparation 
process), where  the middle single-particle layer $\bi C$  forms  
a  defect plane. We did not encounter extrinsic 
 stacking faults with sequence $ABCABC{\bi B}ABC$,  
where the middle layer $\bi B$ is  inserted.

The  disordered structures  in the 
initial particle configurations at $t=0$ 
 have been created  during the crystllization 
 in the preparation process. 
  In Fig.1,   using the initial 
  data of unstrained states, 
  we display   
  relatively disordered  particles 
with  $D_{j}>D_0$   
 in four cases: 
(a) $c=1$ and  $\sigma_2/\sigma_1=1.2$,  
(b) $c=0.95$ and $\sigma_2/\sigma_1=1.2$,  
 (c) $c=0.02$ and  $\sigma_2/\sigma_1=1.4$, 
 and  (d) $c=0.95$ and 
 $\sigma_2/\sigma_1=1.4$.   
The lower bound $D_0$ is 0.05 in  (a) and (d) 
and is  $0.03$  in  (b) and (c), since the particles 
composing intrinsic stacking faults 
have slightly larger $D_j$  than those 
composing  twin  faults.  
Relatively ordered particles with $D_j<D_0$ 
 are transparent in these pictures. 
Then we can visualize stacking faults,  
because  the particles on these planes  
  have higher values of $D_j$ than 
  in the background fcc region.  

In the single component case,  
we  realized  one or two intrinsic 
 stacking faults extending   throughout the cell 
 in many cases, as in (a). We could also realize 
 initial configurations  without stacking faults for $c=1$. 
 In  the upper part of (a), 
  two parallel  stacking faults 
 meet with   separation of  two particle  layers. 
Partial dislocation cores  
are along  the junction line. 
These stacking faults are  parallel  to the $x$ axis 
and make an angle of  $0.30\pi$
with respect to the $z$ axis. 
 In (b), five  twin faults with  single-particle thickness 
can be seen.  
Four of them constitute two pairs of 
parallel  twin faults  
 stemming     from a grain-boundary-like 
disordered region  in the right upper part, while  
 the bottom one is unpaired.  For example, the 
two twin faults  in the left upper part 
are paired as  $AB{\bi C}BA{\bi C}ABC$  with two $\bi C$ 
being visible. 
It is worth noting that 
paired twin faults can easily be produced 
from grain boundaries in nanocrystals  \cite{twin}.

In (c) and (d) of Fig.1, 
the size ratio  $\sigma_2/\sigma_1$ is increased 
from 1.2 to 1.4. See Fig.6 of 
our previous paper \cite{KawasakiJCP} 
for other snapshots  of the particle configurations 
with the same  parameter values, 
where disordered particles with $D_j>0.2$ were shown to  
form tube-like or plate-like 
aggregates.  At $c=0.02$ in (c),    twin  and intrinsic 
stacking faults are mixed, 
where a small number  of 
the large particles strongly disturb the fcc crystalline 
order of the small particles. In (d), 
intrinsic stacking faults 
appear successively to form hcp regions. 
The local composition in 
the hcp regions is about $0.95$, while  
that of the (transparent) fcc regions 
is $0.995$.  
In this case, a small number of  the small particles 
tend to be  expelled from  the 
 fcc and hcp crystal regions 
\cite{KawasakiJCP}.

Here we introduce the surface free   energy 
for   planar defects 
per unit area \cite{Hirth}. 
  In our simulation, it is well-defined  
 in unstrained states with small overall 
 disorder, though  
 its meaning becomes not clear 
 in highly strained states (as will be the case  in Fig.3).   
That is, at low $T$, it is given by 
 the excess potential energy 
 per unit area,  since the entropic contribution 
is small.  Thus     
we  define  the  potential energy 
of  the particle $j \in \alpha$  as
\be 
U_j= \frac{1}{2} 
\sum_{k}v_{\alpha\beta}(|{\bi r}_j-{\bi r}_k|) ,
\en 
where   $k \in \beta$ represent 
the surrounding particles. 
As well as $D_j$, the values of  $U_j$ supported by   
the  particles in  stacking faults 
are   higher than the background fcc value  $U_{\rm fcc}$ 
($\cong -7.40\epsilon$ here). 
For intrinsic stacking faults in the panel (a) in Fig.1, 
the excess potential energy 
$U_j- U_{\rm fcc}$  is about 
  $0.15 \epsilon$ on the two 
 stacking-fault  layers and is 
 about $0.1  \epsilon$  on the next layers. 
  Since  the areal particle  density 
 is about $0.92 \sigma_2^{-2}$,  
we estimate the surface  energy for intrinsic stacking faults as  
\be  
\gamma_{\rm int} \sim  0.5 \epsilon/\sigma_2^2.
\en  
Note that the hcp structure 
is a succession of intrinsic stacking faults as in 
the panel (d) of Fig.1. Therefore, 
the hcp structure should have a higher 
energy than the fcc structure roughly by $0.1\epsilon$ 
per particle for the Lennard-Jones potential. 
On the other hand, for twin  faults in the panel (b), 
$U_j- U_{\rm fcc} $  is about 
  $0.16 \epsilon$ on the  
 stacking-fault  layer   and is 
 about $0.07 \epsilon$  on the next layers.  
 Thus the surface  energy for  twin faults 
 is  estimated as 
\be 
\gamma_{\rm twin}\sim 0.3 \epsilon/\sigma_2^2.
\en  
In our previous paper \cite{KawasakiJCP}, we have estimated the surface 
energy from twin faults in its Fig.3. 

 If we set  $\sigma_2 = 2{\rm \AA}$ 
and $\epsilon =300 k_B$ in Eq.(17),   
we have  $\gamma_{\rm int}\sim 50 $mJ$/m^2$ 
and $\gamma_{\rm twin}\sim 30 $mJ$/m^2$. 
It is worth noting that these  estimated values are 
of the same order as those     for   Cu  and Ag  
\cite{Gallager}. 
In such  nanocrystalline metals,  
  partial dislocations and stacking 
  faults have  been widely  observed.     
Full dislocations are more frequently  observed 
for  metals such as Al and Ni with  larger 
 stacking fault  energies 
 ($\gs  100$mJ$/m^2$)\cite{Yama,Swy}.  
For  hard-core colloids,  
 the stacking fault energy is nearly zero,  
 where coexistence  of fcc and hcp 
structures have been observed  
\cite{Poon,Weiz,De,Blaa,Blaa1,Pusey,Dhont,Chaikin,Solomon}.

\subsection{Elementary stacking-fault motions}

In fcc crystals, an intrinsic   stacking fault  
 appears  with a relative 
displacement ${\bi b}_1$ 
of type $(a/6)\av{{1} {\bar 2} { 1}}$ taking place 
between two adjacent  $\{111\}$ planes\cite{Hirth}. 
The lattice constant   $a$ is close to $2^{2/3}
\sigma_2$ for $c=1$ here.  
A stacking fault disappears  
upon occurrence of  a  second relative 
displacement ${\bi b}_2$ 
of type $(a/6)\av{\bar{2} 1 1}$. The vector sum results  in 
a full Burgers vector  ${\bi b}= {\bi b}_1+{\bi b}_2$ 
of type  $(a/2)\av{\bar{1} 1 0}$.  
In Figs.2-7, we examine the stacking fault dynamics 
in various aspects using a single simulation run for $c=1$.

In Fig.2,  we illustrate an example of succession 
of these two-step processes for $c=1$. 
The first and second  partial displacements occur at 
$t\cong 2000\tau_0$ and $4000\tau_0$ with a duration time of 
order $20\tau_0$. Here  $D_j$ 
on  the stacking fault 
increases after the first slip 
and again decreases after the second slip.     In this case, 
 however, the size of  the first relative displacement 
 depicted in (a) is  $80\%$ of $|{\bi b}_1|=a/\sqrt{2}\cong 0.72 \sigma_2$ 
 because of  the   elastic  stress still 
 remaining below the stacking fault, while 
 the size of the second one 
in (c) is very close to $|{\bi b}_2|=a/\sqrt{2}$.

To examine this slip  dynamics 
 in more detail, 
let us  consider $\{111\}$ layers close to 
the  intrinsic stacking fault treated in Fig.2. 
We pick up  particles in   the $\lambda$-th layer 
below the stacking fault, 
whose positions $(x,y,z)$ are specified by    
 $0< (z+ 0.636y)/\sigma_2 +\lambda-14.5<1$, 
$4<x/\sigma_2<13$, $9<y/\sigma_2<13$, and $3<z/\sigma_2<10$.  
The number of these particles  $N_\lambda$ 
is close to 50.  The stacking fault   itself consists 
of the layers with $\lambda=0$  and $-1$. 
In Fig.2,  we have visualized  the  layers  
with $\lambda=-1,0$, and 1.  Here we 
take the average of 
the potential energy $U_j$ in Eq.(16) 
over these particles to obtain 
\be 
{\bar U}_\lambda(t) = 
\sum_{j \in {\rm layer~\lambda}} U_j(t)/N_\lambda . 
\en  
The left panels of Fig.3 display 
${\bar U}_\lambda(t)$ 
versus $t$  for  $\lambda=0, 3,$ and 6  
at the first and second slips. 
For $\lambda=0$, it exhibits a peak at each  slip, 
where the  peak height 
is   higher by $0.2 \epsilon$ 
than its stating  value 
$\bar{U}_0(t_1)$ or $\bar{U}_0(t_2)$. The resultant 
energy barrier for the (first and second) 
 partial displacements per unit area is estimated as    
\be 
\Delta \gamma_{\rm peak}\sim 0.2\epsilon\sigma_2^{-2}, 
\en 
which is of the same order as $\gamma_{\rm int}$ in Eq.(17) 
 \cite{Swy}. 
Right after the first displacement, 
${\bar U}_\lambda(t)$  drops     
below  $\bar{U}_\lambda(t_1)$ by 
$0.05\epsilon$ for $\lambda=0$  
and by $0.03\epsilon$ for $\lambda=3$, 
while  it   is nearly unchanged for $\lambda=6$. 
The  elastic  energy stored at $t=t_1$  is not  
released  for $\lambda=6$ in this case. 
In contrast, in unstrained states in Fig.1,  
 the particles on  stacking faults 
have higher $U_j$ (by  $0.1  \epsilon$ here) 
than those in  the fcc regions, 
leading to  a well-defined $\gamma_{\rm int}$.   
On the other hand,  the local kinetic energy 
is enhanced only during the burst periods. 
That is,  it  increased  
by $0.05\epsilon$ for  $\lambda=0$ 
and by $0.01\epsilon$ for $\lambda=6$   per particle  
right after the two displacements,  
but the kinetic-energy  peaks  rapidly decayed 
  on a time scale of $20\tau_0$ (see $\bar{K}(t)$ in Fig.5 
  and the remark (1) in the last section).

The right panels of Fig.3 display 
 the time-dependent displacement 
of the $\lambda$-th layer defined by 
\be 
\Delta {\bar r}_\lambda(t)=\sum_{j\in {\rm layer~\lambda}}
|{\bi r}_j(t)-{\bi r}_j(t_0)|/N_\lambda,  
\en 
where $t_0=t_1=1980\tau_0$ or 
$t_2=4000\tau_0$ as in Fig.3 
and  the average is taken  over 
the $N_\lambda$ particles picked out in Eq.(19). 
In  the first slip, 
the displacement  at $\lambda=-1$ is opposite 
to that at $\lambda=0$ with a nearly equal magnitude. 
In  the second  slip, 
 the displacement at $\lambda=0$ 
is over a distance of $0.38\sigma_2$,  
while that at $\lambda=-1$ 
makes  an angle of $3\pi/4$ 
with respect to that at $\lambda=0$ 
with  a  magnitude of  $0.45\sigma_2$. 
Then the   relative displacement  
vector is nearly equal to ${\bi b}_2=
 (a/6)[\bar{2} 1 1]$ as in the panel (c) of Fig.2.  
The maximum of the sliding velocity of the 
layer $\lambda=0$ is about $0.12\sigma_2/\tau_0$ 
both for the first and second partial  displacements. 
 so the maximum of the relative velocity between the two layers 
is $0.24\sigma_2/\tau_0$. 
We shall see that the propagation 
velocity of  a  stacking fault  edge 
 is about $4\sigma_2/\tau_0$ in Fig.7.

In the left panels in Fig.3,  
$\bar{U}_0 (t)$ is lowest 
right after the first slip 
but is largest right before the second slip among 
  $\lambda=0,3,$ and 6, 
   owing to  inhomogeneous 
accumulation of the elastic energy. 
In  Fig.4, we thus presents 
time-evolution of $\bar{U}_\lambda (t)$ 
on a long time interval in  the same simulation run. 
In addition to the two slips in Figs.2 and 3,  we notice 
occurrence of another slip 
 at $t\sim 2500\tau_0$  
  away  from the layer elements 
under consideration. Remarkably, 
at this remote slip,  the elastic energy density  
  decreases  suddenly by $0.05\epsilon$ 
for  $\lambda=3$ and 6, but it 
is  unchanged for  $\lambda=0$, 
resulting in $\bar{U}_0 (t)> \bar{U}_3 (t)>\bar{U}_6 (t)$ 
for $t>2500\tau_0$.  
  In the inset of Fig.4, 
 we also display  the average disorder variable 
 $\bar{D}_\lambda(t)$ defined by  
 \be 
{\bar D}_\lambda(t) = 
\sum_{j \in {\rm layer~\lambda}} D_j(t)/N_\lambda .  
\en  
For $\lambda=0$, ${\bar D}_\lambda(t) $ 
  increases  at $t\cong  2000\tau_0$ 
and  decreases  at $t\cong  4000\tau_0$, 
while it remains to be the background fcc value 
for  $\lambda=3$ and 6.

\begin{figure}[t]
\includegraphics[scale=0.5]{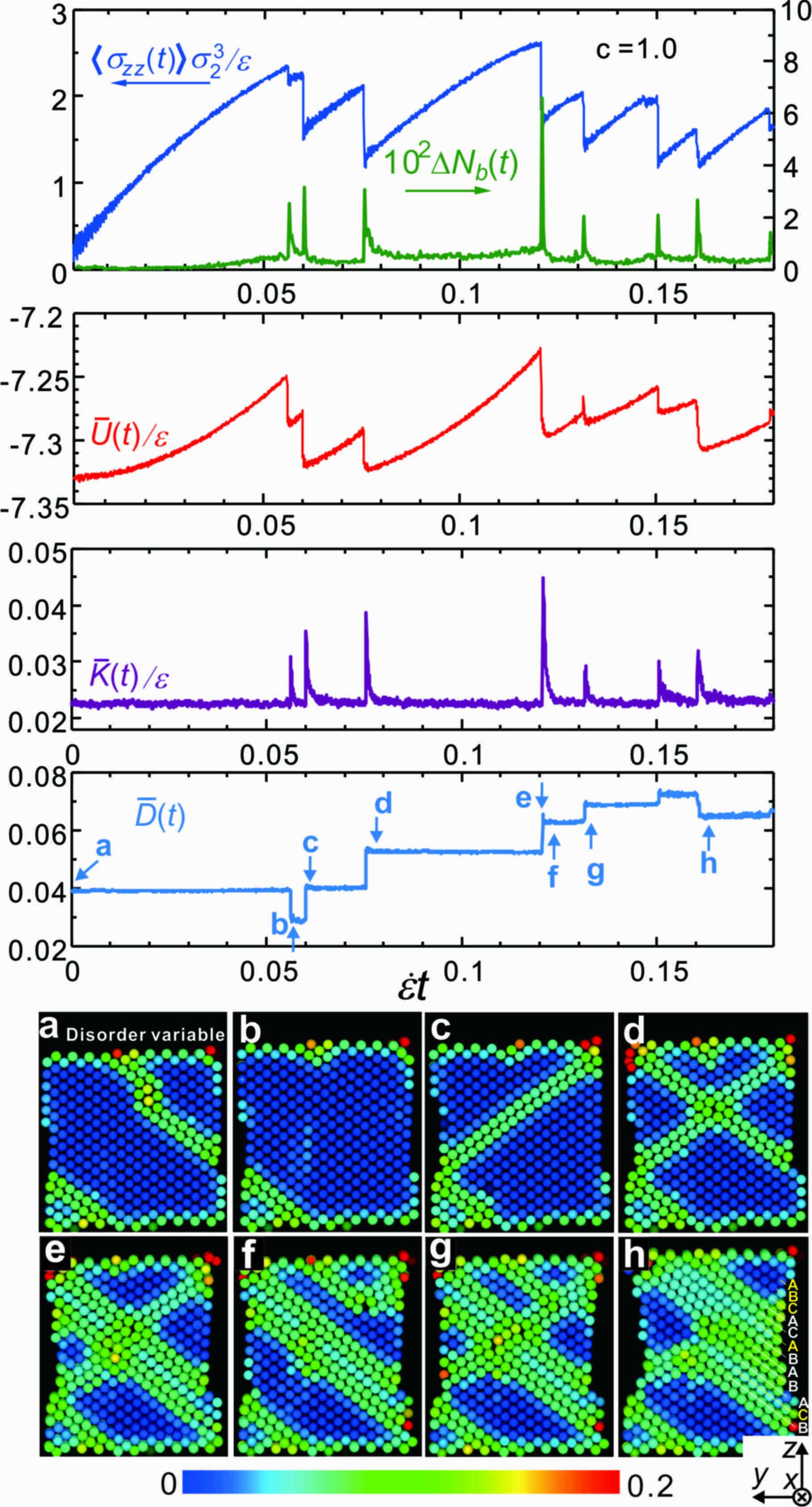}
\caption{(Color online) 
Time evolution of averages 
over the unbound particles for $c=1$ under  stretching: 
$\av{\sigma_{zz}}(t)$, 
$\bar{U}(t)$, $\bar{K}(t)$, and ${\bar D}(t)$ 
from above. Broken bond numbers $\Delta N_b (t)$ 
are also shown (top),  where  $\Delta t=10 \tau_0  =
0.0003/\dot{\varepsilon}$ and $A_2=1.31$. 
Drops  of $\av{\sigma_{zz}}(t)$  and 
$\bar{U}(t)$  and bursts of $\Delta N_b (t)$  
 and $\bar{K}(t)$  occur  simultaneously 
 at slip events.   
Bottom: $D_j(t)$ in a plane ($\perp$ the $x$ axis) 
at eight points (a)- (h) 
marked on the curve of ${\bar D}(t)$, 
where stacking faults are all intrinsic ones.  
Initial one in (a) disappears in (b), 
while new ones emerge in (c), (d),  and (e). 
In (f) one disappears at $ t \sim 4000 \tau_0$, which  
corresponds to the second partial displacement in Figs.2 and 3.  
In (h) a rhcp region is realized 
with sequence $ABCACABABACB$.}
\end{figure}

\begin{figure}[t]
\includegraphics[scale=0.42]{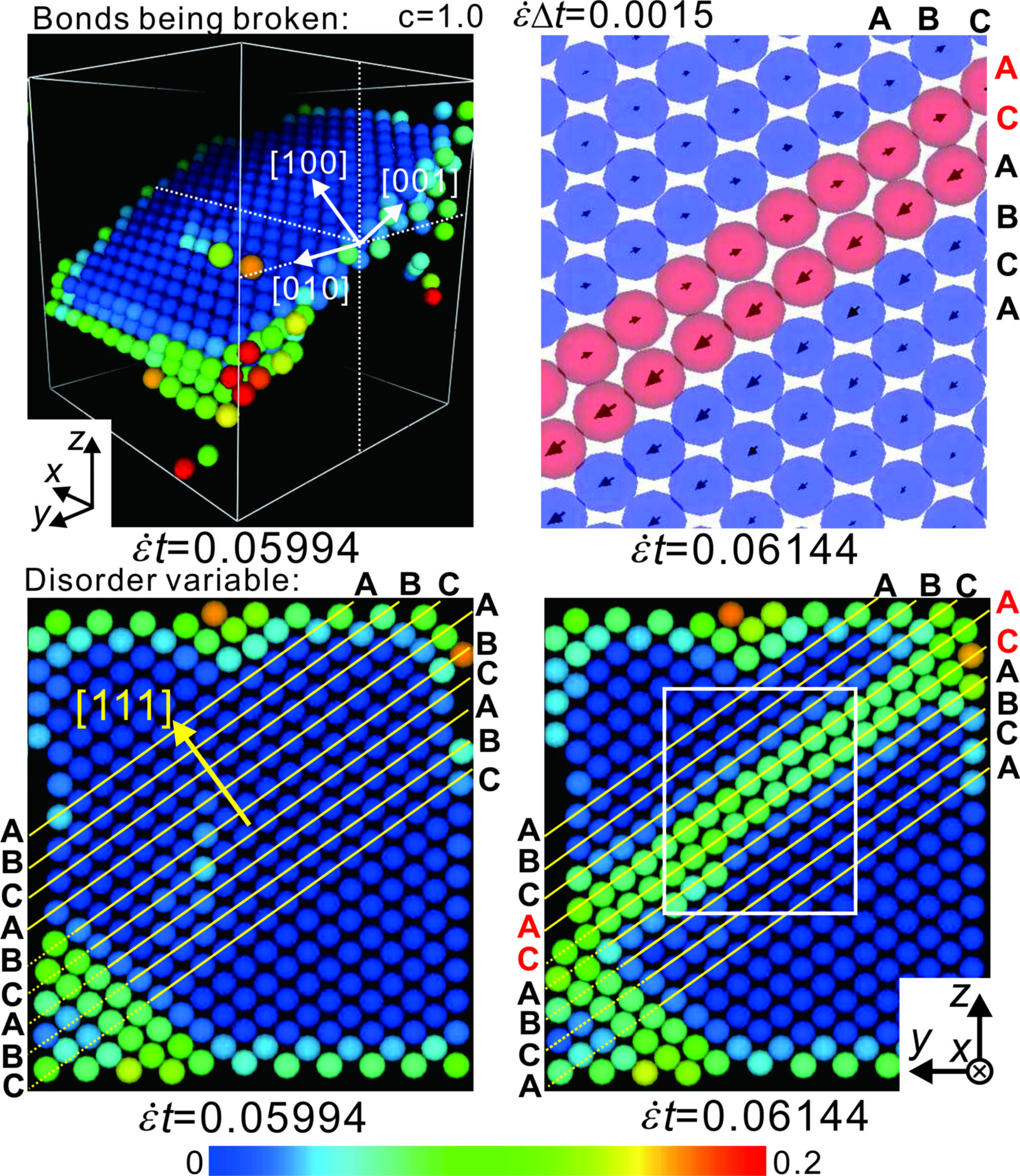}
\caption{(Color online)    
Formation  of an intrinsic   stacking fault 
as a slip for  $c=1$.    In the left top panel,  
particles are displayed at ${\dot \ve} t =  0.05994$, 
whose bonds  are  broken  
  subsequently  in  
 $0<{\dot \ve} t -  0.05994<{\dot \ve}\Delta t$. 
 We set  $ \Delta t= 50\tau_0$,  $A_1=1.3$,  and $A_2=1.4$.
Three fcc principal axes are written (in white). 
Other panels show particles     
in the region $ 9.93<x/\sigma_2<11.03$ on a plane ($\perp$ the $x$ axis). 
Right top: Those after the slip  around the 
two-particle  layer (in red) 
with large opposite 
displacements, 
where the  white bordered region in the right  bottom 
panel is enlarged.  
Bottom: Those  at 
${\dot \ve} t=0.05994$ (left) 
and  at ${\dot \ve} t=0.06144$ (right) 
 before and after the stacking fault   formation, 
 respectively.  Colors in the left top and bottom represent 
$D_j(t)$ according to the color bar. 
}
\end{figure}


\begin{figure*}
\includegraphics[scale=0.8]{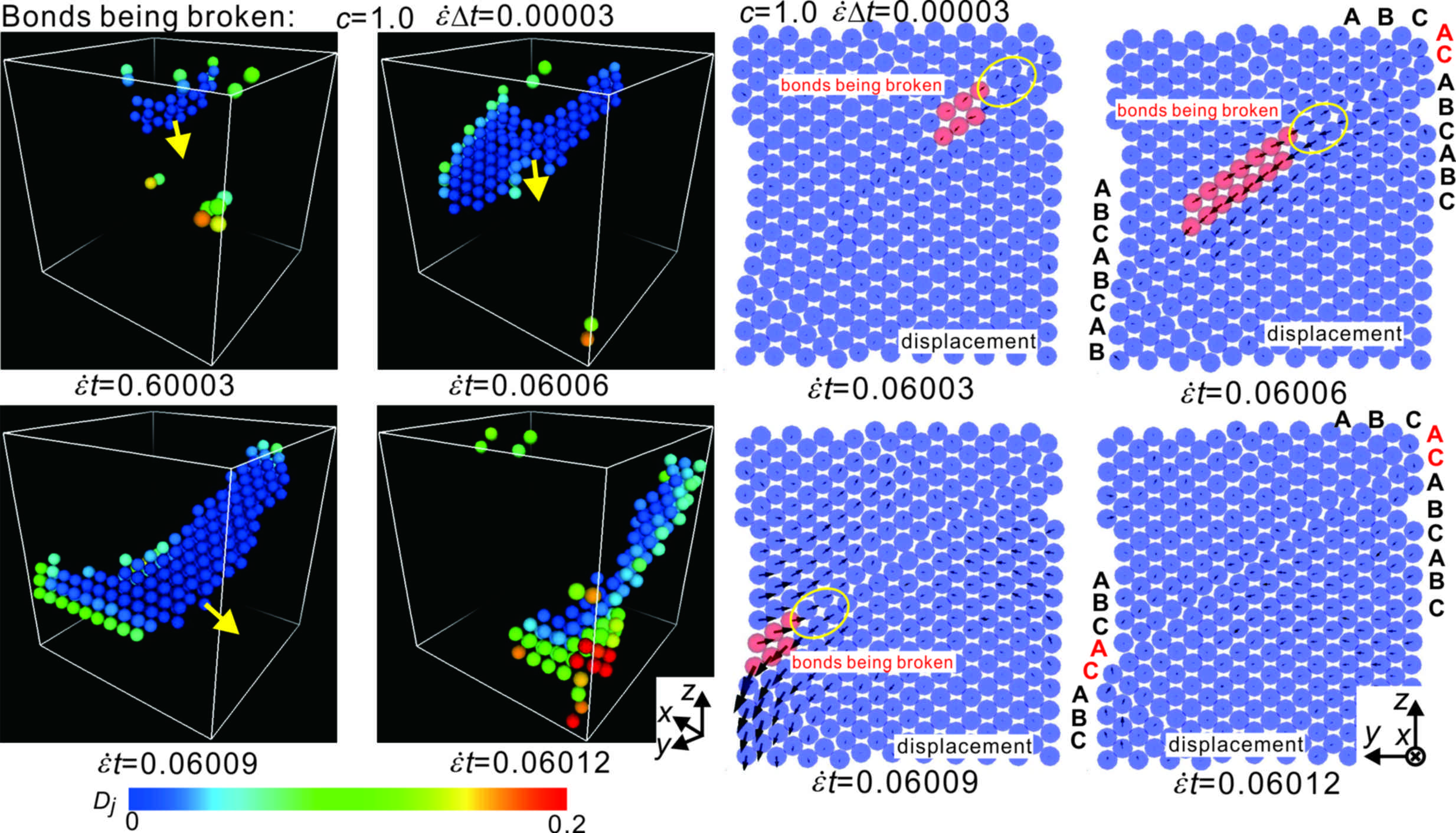}
\caption{(Color online) Birth and growth  
of an intrinsic  stacking fault for $c=1$. 
Left four panels: Particles with their bonds being 
broken    in successive time intervals 
 $n\dot{\ve }\Delta t< 
 \dot{\varepsilon }t- 0.60003<(n+1)\dot{\varepsilon }\Delta t
 $ ($n=0,1,2,3)$ 
 with  $\Delta t=\tau_0$,  $A_1=1.3$,  
 and $A_2=1.31$. Colors represent  
 $D_{j}$ according to  the color bar. 
Arrows (in yellow) indicate  
the direction of the fastest expansion 
of the partial dislocation curve. 
Right  four panels: Particles on a plane 
in the range $12.4 < x/\sigma_2 <13.5$ with displacements 
multiplied by four. 
Red ones are those with their bonds broken 
in each time interval  with $A_1=1.3$ and $A_2=1.31$. 
The edge  velocity   is $4 \sigma_2/\tau_0$.  
The partial dislocation core 
is located behind the edge  within the ellipse in each panel. 
}
\end{figure*}



\subsection{Stacking faults as slip elements }

In Fig.5, we follow time-evolution  of 
 four averaged  quantities 
 during stretching in the strain range 
 $0<\dot{\ve}t<0.2$  for $c=1$, where the initial 
 state is  in the panel (a) of Fig.1.  
 In its  top panel, we plot  
  the average stress 
 $\av{\sigma_{zz}}(t)$ in Eq.(15)  and  
 the broken bond numbers 
 $\Delta N_b(t)$ defined by Eqs.(10) and (11) 
 with $\Delta t=10\tau_0$.  
   In the initial  stage $0< \dot{\ve}t\ls 0.02$, 
it approximately holds the linear elastic relation,  
\be 
\av{\sigma_{zz}}(t) = E \dot{\ve}t. 
\en 
where the coefficient  $E$ is  
 about $60 \epsilon/\sigma_2^3$.    
Though the top and bottom boundaries are clamped here, 
$E$ is close to Young's modulus 
$3GK/(K+ G/3)\sim 3G$, where  the shear modulus $G$ 
is about  $20 \epsilon/\sigma_2^3$ and  the bulk 
modulus is  about  $100 \epsilon/\sigma_2^3$ for $c=1$. 
At  each slip,  
$\av{\sigma_{zz}}(t)$ suddenly decreases 
and $\Delta N_b(t)$  exhibits a sharp peak  
with heights  mostly about 300 
$(\sim (L_0/\sigma_2)^2$). 
The maximum in this run is about 700 at $\dot{\ve}t=1.2$. 
These  bursts last   on a time scale of order 
$10\tau_0$. The resultant plastic strain is of order 
$\Delta N_b(t)/N_{\rm ub}$ 
and the typical stress drop is estimated as 
\be 
\Delta\sigma\sim G\Delta N_b(t)/N_{\rm ub}\sim 
 \epsilon/\sigma_2^3, 
\en  
which is consistent with   the top panel of Fig.5.

 The two middle panels of Fig.5 
 give   the potential 
and kinetic energies averaged over the unbound particles, 
\bea 
{\bar U}(t)  &=& {\sum_{jk \in {\rm ub}}}'
{v_{\alpha\beta}(r_{jk})}/{2N'_{\rm ub}}, \\ 
{\bar K}(t)  &= & \sum_{j \in {\rm ub}}
{m_\alpha|{\dot{\bf r}_j} |^2}/{2N_{\rm ub}}.   
\ena 
In  defining   the average potential, 
we should note that the particles 
on the free side boundaries 
have potential energies higher than 
those in the interior by $3-5 \epsilon$. 
Thus, in the summation of  ${\bar U}(t)$,  
we have removed   the particles near the side 
boundaries with distances shorter than 
$3\sigma_2$. The number 
$N'_{\rm ub}$ in Eq.(25) is then that 
 of the unbound particles 
 with larger separation from the side boundaries. 
  After a slip  at $t=t_0$, 
 ${\bar U}(t)$ grows roughly  as 
\be 
\Delta {\bar U}(t) =
{\bar U}(t)- {\bar U}(t_0) 
\sim   E{\dot\ve}^2(t-t_0)^2/2.
\en 
At  each slip,  a large fraction of the 
elastic energy  is eventually absorbed by  
the thermostats at the top and bottom  very effectively  because of 
short $\tau_{\rm NH}$ and the small system size in our 
simulation.   However,  it  
 is partly  used to create the new stacking fault. 
Formation of a  stacking fault 
with area $L_0^2$  yields  an increase in ${\bar U}(t)$ 
of order,  
\be 
\Delta{\bar U}_{\rm sf} 
\sim 
\gamma_{\rm sf}L_0^2/N_{\rm ub}  \sim 0.02\epsilon. 
\en    
On the other hand,  
 the average kinetic energy    $\bar{K}(t)$ exhibits 
  sharp peaks smaller than $ 0.04\epsilon$  
 decaying  to zero rapidly, so   $\bar{K}(t) \cong 
 3k_BT/2\cong 0.0225\epsilon$ holds 
 except for  the burst periods. 
 Thus  heating is negligible. 
Furthermore,   ${\bar D}(t)$ in Fig.5  
increases or decrease 
in a stepwise manner upon appearance or disappearance 
of a stacking fault.  
This coincidence 
can be seen from   side views of stacking faults 
in  the eight bottom plates (a)-(h) in   Fig.5. 
Between (e) and (h) 
a stacking fault disappears, which corresponds to 
the second partial displacement in Figs.2 and 3. 
In  (h) we can see formation of a 
random sequence of stacking (a random hcp region).

In  Fig.6, we illustrate a slip event   for  $c=1$,  
where an intrinsic $(111)$   stacking fault 
is created in a time interval 
 $0<{\dot \ve} t - 
 0.05994<{\dot \ve}\Delta t= 0.0015$. We set 
 $\Delta t=50\tau_0$,  $A_1=1.3$,  and $A_2=1.4$. 
The system length is increased 
only  by $0.02\sigma_2$ in this   time interval. 
In the left top, the displayed particles  are just before  
the slip, so they still have the background 
fcc value of $D_j$ except for those 
near the free boundaries. Here the  crystal orientation 
can be known  from    the crystal  principal axes depicted, 
where  $[010]$  is parallel to 
the $y$ axis, while 
$[001]$ and and $[100]$ 
 make angles of  $\pm \pi/4$ 
with respect to the $z$ axis in  the $xz$ plane. 
 Thus the slip plane  
  makes  an angle of $\cos^{-1}(1/\sqrt{3})= 0.30 \pi$ 
with respect to   the free surfaces (parallel to  
the $xz$ plane). The other panels 
give cross-sectional particle configurations 
in the region $ 9.93<x/\sigma_2<11.03$. 
The right top panel displays 
the  layer stacking and the displacements 
at the slip. The bottom panels 
show that  $D_j$  change 
on the stacking fault plane  after 
the slip event.

The formation of a stacking fault 
takes place in a very short time ($\ls 10\tau_0$) 
in our small cell. 
Thus, in  Fig.7, we  give  sequences of 
snapshots of its growth for $c=1$, 
where the time width is 
taken to be $\Delta t=\tau_0$. 
In each panel in the left, particles 
 with their bonds being broken 
are displayed, where we set 
 $A_1=1.3$ and  $A_2=1.31$. 
The direction of the fastest expansion 
of the stacking fault is marked by arrows. 
The   expansion  velocity  of the edge 
 is of order $4 \sigma_2/\tau_0$,  
which is slightly smaller 
than the transverse sound velocity  
$c_\perp =(G/m_2 n)^{1/2} \sim  5 \sigma_2/\tau_0$. 
Note that the actual particle  velocity 
undergoing a slip is of order 
$ 0.1 \sigma_2/\tau_0)$,  
as  already demonstrated in Fig3,  and 
is much slower than the  edge-expansion velocity. 
Thus the values of $D_j$ of the  displayed 
particles (except those near the free boundaries) 
are still  close to their background fcc value. 
In the four right panels,  we visualize cross-sectional 
 particle configurations around the  expanding stacking fault, 
where it  is  at its birth in the upper left panel, expands through the cell, 
 and colloids with the free boundary 
in the bottom panels. 
Here, because $A_2$ is only slightly larger than 
 $A_1$,  the edge  represents   an inception of 
the bond breakage or a precursor of 
the partial dislocation. The relative displacement 
approaches its saturation length 
$a/\sqrt{6}$ at  the partial dislocation within the ellipses 
in the right panels in Fig.7. The particles within 
these ellipses mostly have 11 bonds in 
our  definition in Eq.(10). 
This precursor zone increases  in time to have  a width of 
4-8 particles in the present  case.   
In addition, we note that 
the instantaneous particle velocities $\dot{\bi r}_j$ 
around the edge are rather random  with 
large deviations   because of 
their rattling motions in the crystal structure. 
Their time averages over $\tau_0$ 
given by $ 
\int_t^{t+\tau_0} dt' \dot{\bi r}_j(t')/\tau_0$ 
become   proportional 
 to the displacement vectors in the panels.

We finally  examine   
how  stacking faults are removed 
by a collective partial displacement 
in applied strain. We start  with the initial configuration (b) 
in Fig.1 for   $c=0.95$, where the initial stacking faults 
were twin ones but an intrinsic stacking fault 
was  created subsequently. Here the degree of disorder 
is considerably  larger than in the case of $c=1$.  
 The left top panel of Fig.8 
 displays  particles whose bonds 
 are just before breakage.  
 They are mostly on the disappearing 
 intrinsic stacking fault plane,  
  but those  in the projected part  are 
  in the grain boundary-like region 
  in the upper part in the panel (b) 
  of Fig.1. They are   thus relatively disordered before the slip. 
The right top panel gives 
a cross-sectional profile of the stacking fault
just before its disappearance, where 
 the displacement vectors are 
 from the present to  future positions  
 after a time elapse of $\Delta t=50\tau_0$. 
In the bottom plates, the values of 
the disorder variable are displayed 
in the plane before and after the second partial 
displacement. 
The stacking  $CBCA$ ($\perp$ $[111]$) in the left  
disappears in the right 
by the registry change: 
$B\to A$, $C\to B$, and $A\to C$ 
as  marked in the left bottom corners.

\begin{figure}[t!]
\includegraphics[scale=0.5]{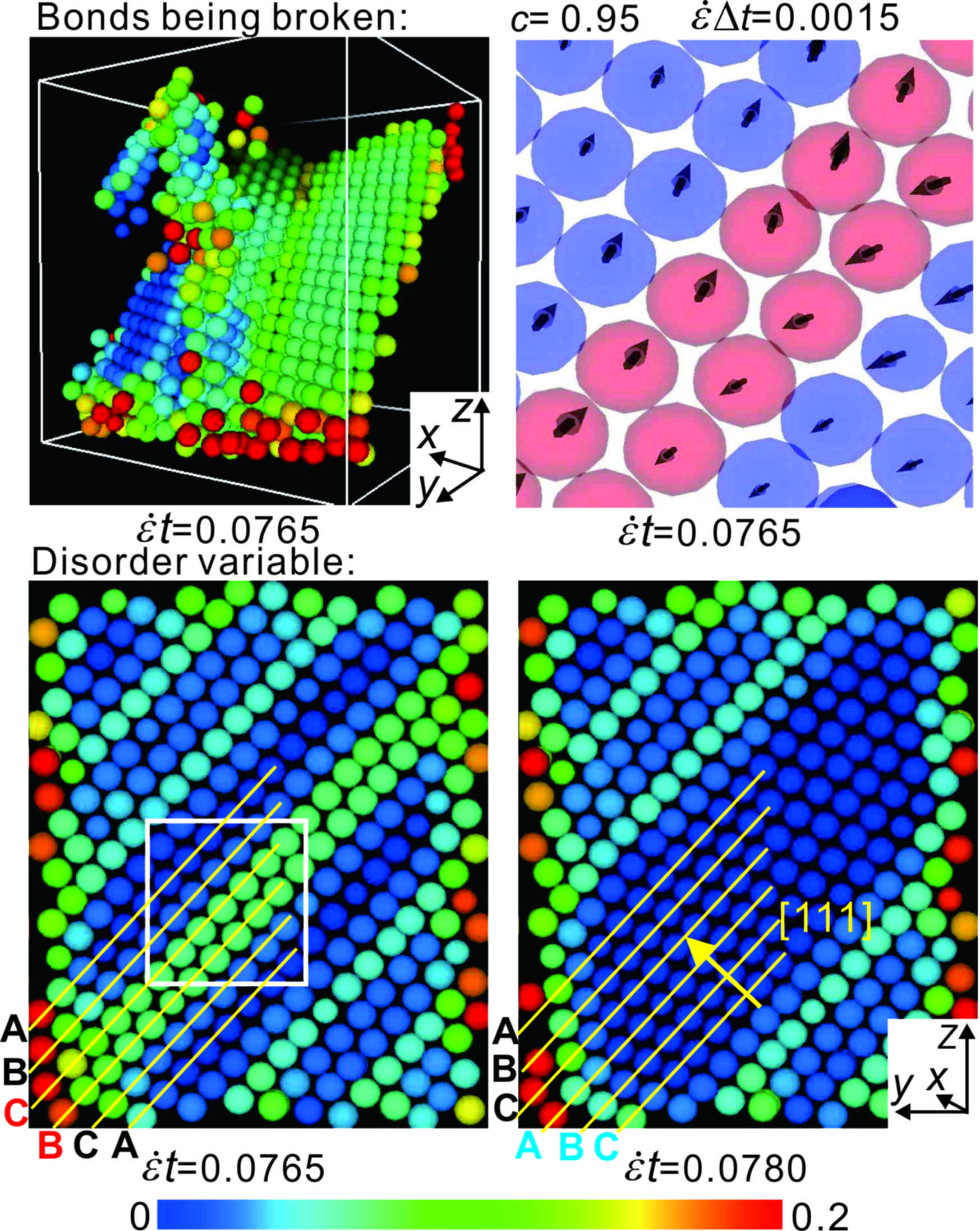}
\caption{(Color online) 
Removal of  a preexisting intrinsic  
stacking fault  by a subsequent 
 partial displacement   
 for  $c=0.95$. Colors  represent $D_{j}$ according to the color bar.
Top left: Particles whose bonds are  just before breakage 
at the second  displacement, so they  
mostly have relatively large $D_j$ before the slip.  
Top right: Particles in a plane 
in the range $3.9<x/\sigma_2<5.0$. Arrows   represent  
displacements multiplied by 1.5   
in a  time interval with width $\Delta t=50\tau$.
Red particles have broken bonds 
with  $A_2=1.35$.
Bottom: $D_j$ of particles on the plane  
before the slip at $\dot{\varepsilon}t=0.0765$ (left) 
and after the slip at $\dot{\varepsilon}t=
0.0780$ (right).
}
\end{figure}

\section{Summary and remarks } 

In Lennard-Jones particle systems, 
we have studied  the 
dynamics of stacking faults  
with the aid of the bond breakage, 
  the  disorder 
variable $D_j(t)$, and the particle displacements 
at $T=0.015\epsilon/k_B$. 
Hereafter,  we summarize our main  results 
together with comments.\\
(i) We can select relatively disordered  particles 
with $D_j$ larger  than a threshold $D_0$.  For appropriate $D_0$,  
we can visualize   stacking faults because the particles 
 belonging to them have higher values of $D_j$ 
 than those in the fcc crystal region.  
 In Fig.1, we have shown the initial 
structural disorder  for four cases 
of  the composition $c$  and the size ratio $\sigma_2/\sigma_1$   
to find intrinsic  stacking faults, twin faults,  
and hcp regions.  They indicate emergence of 
a vast variety of  
of the structural disorder sensitively  
depending on $c$  and $\sigma_2/\sigma_1$.
\\
(ii) 
We have shown that 
plastic deformations in fcc crystal can be  achieved 
by partial displacements of 
close-packed planes, as illustrated in Fig.2. 
We have examined rapid time-development 
of the potential energies  and the positions    
of the particles on a stacking fault and 
those on layers nearby at the first 
and second partial displacements in Fig.3. 
The potential energy of the particles 
 increases by $0.2\epsilon$ 
 when they are on an emerging or disappearing 
 stacking fault. The relative particle velocity 
at stacking faults is of order $0.24\sigma_2/\tau_0$ 
and is much slower than the transverse sound velocity 
$c_\perp\sim 5\sigma_2/\tau_0$. 
We have detected 
the position-dependent 
elastic energy stored during  
the elastic periods  and  its collective release 
upon slips   in Fig.4.  
\\
(iii) We have shown time-evolution of 
some physical quantities over long times 
during stretching in Fig.5.
Upon appearance and disappearance 
of stacking faults, the  stress 
exhibits strong intermittent 
fluctuations with collective bond breakage. 
The average elastic energy 
accumulated in the cell  is suddenly  decreased  upon slips.\\
(iv) Apperance of an intrinsic stacking fault 
has been illustrated in Fig.6.  
The formation of stacking faults 
occurs very rapidly, so Fig.7 
has given  the time-evolution of a stacking fault 
on a time scale of $\tau_0$. 
Remarkably, the edge of the 
stacking fault expands  
with a velocity close to the transverse sound 
velocity $c_\perp$. As a result,  there appears 
a precursor zone between the edge 
and the partial dislocation, 
whose width increases in time and 
is in a range of 7-10 particles in our case. 
Disappearnce of an intrinsic stacking fault 
has been illustrated in Fig.8, where 
intrinsic stacking faults proliferate 
to form a rhcp sequence 
on a long time scale, however. 
\\

We further make critical remarks as follows:\\
(1)  Because of  the choice $A_1=1.3$ 
in the definition of bonds in Eq.(10), 
 we have picked up the 12 nearest neighbors 
 as bonded particles in fcc crystal.  
The disorder variable  $D_j$ in Eq.(13) 
has then aquired the meaning of the deviation of hexagonal order 
in the fcc crystal structure.\\  
(2) In our simulation, we have realized intrinsic 
stacking faults during stretching. 
However, in previous experiments on nanocrystal plasticity \cite{twin}, 
paired twin faults like  
$BACB{\bi A}BCABC{\bi A}CB$ have been observed 
to appear from grain boundaries, 
where  the second and last ${\bi A}$ 
form twin faults. \\ 
(3) This paper has treated 
elementary   dynamics of 
stacking faults in a small system with 
free boundaries.  Thus 
stacking faults have appeared 
 from the free side boundaries.  Larger  system sizes 
are needed  to examine   large-scale 
plastic events. In particular, 
we should examine plastic deformations 
in polycrystals with many grains, 
where birth and growth of partial and  full 
dislocations are strongly influenced by 
the grain boundaries.  
With decreasing the grain size, the grain-boundary 
sliding should also come into play.\\  
(4) As indicated by Fig.1, 
the structural disorder 
becomes very complicated 
with increasing the composition 
of the second component. 
This disordering effect is much 
intensified for larger size ratios 
between the two particle species. 
Thus we  should study the effects 
of increasing disorder in binary mixtures 
with size dispersity.\\  
(5) To understand the slip behavior in 
nanocrystals, Swygenhoven {\it et al.}  
\cite{Swy} proposed to use a generalized stacking fault energy 
as a function of the particle displacement, 
because there are potential barriers  for 
the formation of partials. From  the left panels 
of Fig.3, we recognize  that the generalized stacking fault energy 
should furthermore depend 
 on the local elastic strain.\\ 
(6) We have attached Nos\'e-Hoover  thermostats 
at the top and bottom layers in the cell 
and have applied stretching by pulling 
the particles in the top layer. 
We should examine  this method in more detail 
 by varying the parameters such as 
the characteristic time 
$\tau_{\rm NH}$ and the strain rate $\dot{\ve}$. 
In this paper,  $\tau_{\rm NH}$ 
is  very small $(=0.072\tau_0$) 
and the thermal equilibration after 
plastic events is very fast ($\sim 20\tau_0$). 
However, for longer   $\tau_{\rm NH}$, 
the thermostats become less efficient 
and the bulk thermal relaxation  becomes longer,  
resulting in heating in  the cell. 
In larger systems, 
plastically deformed regions should emit 
sound  waves, which  propagate throughout 
the solid region and are reflected 
at the boundaries \cite{dislocation}.  This acoustic effect should 
 be studied in future.

\begin{acknowledgments}
We would like to thank 
 Hayato Shiba, Toshiyuki Koyama,  
 and Tomotsugu  Shimokawa 
 for informative discussions. 
This work was supported by  Grant-in-Aid 
for Scientific Research 
 from the Ministry of Education, 
Culture, Sports, Science and Technology of Japan. 
One of the authors (T. K.) 
 was supported by the Japan Society for Promotion of Science.
\end{acknowledgments}


\begin{references}
\bibitem{stress-comment} 
 The stress components 
 measurable in experiments are those exerted 
 on a  boundary wall. These boundary values  
  coincide  with  the space-averages 
  on time scales longer than 
 the acoustic time $L_0/c_\perp$ with 
 $c_\perp $ being the transverse 
 sound velocity, as confirmed in 2D \cite{Shiba}.
\end{references}

\end{document}